\documentclass{optica-article}

\journal{opticajournal} % for journals or Optica Open

\articletype{Research Article}
\usepackage{float}
\usepackage{lineno}
\usepackage{amsmath}
\begin{document}

\title{Extraction of power transmission parameters from PT-symmetric waveguides}

\author{Chengnian Huang,\authormark{1,3} Zhihao Lan,\authormark{2} M englin L. N. Chen, \authormark{3,4} and Wei E. I. Sha\authormark{1,5}}

\address{\authormark{1}College of Information Science Electronic Engineering, Zhejiang University, Hangzhou 310027, China\\
\authormark{2}Department of Electronic and Electrical Engineering, University College London, Torrington Place, London WC1E 7JE, United Kingdom\\
\authormark{3}Department of Electrical and Electronic Engineering, The Hong Kong Polytechnic University, Kowloon, Hong Kong, China}

\email{\authormark{4}menglin.chen@polyu.edu.hk}
\email{\authormark{5}weisha@zju.edu.cn}

% use {asbstract*} to suppress the copyright line. Copyright information will be added in production

\begin{abstract*} 
The PT-symmetric waveguides have been frequently discussed in the photonics community due to their extraordinary properties. Especially, the study of power transmission is significant for switching applications. The aim of this study is to extract the mode power transmission parameters based on the coupled mode equations and analyze the power properties of the PT-symmetric system. The equations relying on the coupled mode theory are constructed according to the two different orthogonality relations between the original and adjoint system. The results matching well with the finite difference simulations demonstrate the validity of our method, while the conventional coupled mode theory fails. The power properties in the PT-symmetric and PT-broken phases are also observed. Furthermore, a new integration is implemented from which the conserved quantity is defined and extracted, which reflects the Hamiltonian invariant of the system.
Our method fully incorporates the properties of complex modes and allows the study of the power transmission properties based on the orthogonality relations, which is also applicable to other types of non-Hermitian optical systems. This work provides a new perspective for the power analysis of PT-symmetric waveguides and is helpful to design the switching devices.
\end{abstract*}

%%%%%%%%%%%%%%%%%%%%%%%%%%  body  %%%%%%%%%%%%%%%%%%%%%%%%%%
\section{Introduction}
The coupled mode theory (CMT)\cite{S.A.Schelkunoff} is a well-established tool of analyzing and designing waveguides, which can be employed to calculate the propagation constant, eigen-frequencies and transmission properties of different waveguide structures. It describes the orthogonality relation of regular waveguide modes and the interactions between different modes of irregular waveguides, which is a productive approach to model the parametric nonlinear devices such as waveguide structures, optical fiber networks, photonic crystal waveguide and various other optoelectronic structures \cite{H.A.Haus,peall1989comparison,W.H.Louisell,haus1999theoretical,xu2000scattering,fan2003temporal}. The conventional CMT (CCMT) has been further developed by many authors\cite{D.Marcuse,snyder1972coupled
,yariv1973coupled,chuang1987coupled,haus1987coupled} and enriched by the non-orthogonal coupled mode theory (NCMT)\cite{haus1987coupled,chuang1987coupled,vassallo1988couplednct,streifer1987reformulationnct,hardy1985couplednct,suh2004temporalnct,huang2009complexnct,bellanca2018assessmentnct}. In recent years, the parity-time (PT) symmetry of photonic structures has been a subject of intensive investigation\cite{alaeian2014nonPT,el2007theoryPT,lan2023large,feng2013experimentalPT,hodaei2014parityPT,lin2011unidirectionalPT,ruter2010observationPT,shi2015limitationsPT,zhu2014pt,yang2023electromagnetically,Pai-Yen2020pt,zheng2019singular}. The quantities of power propagation in individual channel waveguides are of critical importance for switching applications, especially in presence of loss and gain\cite{lupu2013switching,vctyroky2010waveguide}. The construction of coupled mode equation relying on the CMT is an indispensable tool for extracting the power transmission. For the lossless system, the CCMT constructs the mode equations of the the Hermitian system with conserved quantity according to the complex conjugate inner-product of the reciprocity theorem, while for the PT-symmetric optical structures of the non-Hermitian system, the traditional conjugate inner-product fails. Regarding this lossy issue, the NCMT proposed by Hardy \cite{hardy1985couplednct} encompasses refractive index perturbations not included within the waveguides. The Lagrangian treatment developed in Ref.\cite{el2007theoryPT} is used for the problems with real propagation constants of the dual PT-symmetric structures. Nevertheless, the general CMT proposed by the \cite{xu2015general} effectively solves the dispersion relation through the introduction of the adjoint system. It helps to design new optical devices and study the power transmission properties of more complicated structures. Moreover, the PT-symmetric structures of the non-Hermitian system exhibit intriguing optical phenomena, e.g., a real-to-complex spectral phase transition\cite{Nolting} from the PT-symmetric phase to the broken phase, it is worthwhile to analyze the corresponding power properties and the orthogonality relations of the system mode.   

In this work, we derive two kinds of coupled mode equations for the PT-symmetric waveguides based on the different mode orthogonality relations between the original and the adjoint systems\cite{chen2019generalized}. The \cite{xu2015general} considers the adjoint fields under the chiral symmetry system and uses the perturbation method to establish the inner-product equation. On this basis, we additionally discuss the conjugate case and analyze the mode power transmission properties of the individual waveguide under the weak coupling conditions. Numerically, we employ the 3D finite-difference frequency-domain (FDFD) method\cite{G.Veronis,W.Shin,yee1966numerical} to simulate the PT-symmetric waveguides and compare the results of lossy and gain mode powers calculated by the CMT and the FDFD method, which show excellent agreement. Interestingly, before and after the phase transition, the mode power exhibits different transmission properties and non-conserved behaviors, which are attributed to the dispersion properties and the non-Hermitian Hamiltonians of the system\cite{suchkov2016nonlinear}. Also, the orthogonality of the system mode powers has altered accordingly. Despite the non-Hermitian nature of the Hamiltonian, the symmetry refers to invariance under parity (P) and time reversal (T) operations reflecting the Hamiltonian invariant of the system, which allows the system to exhibit other conserved quantities. We demonstrate that the new integration between the supermodes and the total fields  results in a conserved quantity Q within the real spectrum. The symmetric periodic oscillatory curves of ``lossy” and ``gain” mode quantities of Q can be extracted based on the composite operations of supermodes. Finally, we find that the conserved quantity exists only in this PT-symmetric system, which does not manifest in other lossy systems. Our work may contribute to the study of mode power transmissions in more complicated systems. Also, it provides a new perspective on studying and extracting the conserved quantity of the PT-symmetric systems.

The remainder of the paper is organized as follows. In Section 2, we derive the reciprocity theorem and get two different mode orthogonality relations of the complex mode fields between the different systems. The corresponding coupled mode equations are then constructed. In Section 3, we study the mode dispersion of the PT-symmetric waveguides. The transmission parameters of the lossless and the PT-symmetric waveguides are compared with the theoretical results, also with those from the CCMT. Moreover, the conserved quantities are defined and calculated. In Section 4, we summarize the paper.

\section{Basic Theory}
\subsection{The symmetry properties of operators and reciprocity theorem}
In general, a linear operator can be written symbolically as
$Gf\rangle=h\rangle$, and the inner product between two vectors $f(x)$ and $g(x)$ over the domain $a < x < b$ can be defined as $\langle f, g\rangle=\int_{a}^{b} d x f(x) g(x)$. For the scalar inner-product, the transpose of the operator $G$, denoted as $G^T$, is an operator such that $\langle f, G g\rangle=\left\langle g, G^T f\right\rangle$. Hence, the operator $G$ is symmetric if $\langle f, G g\rangle=\langle g, G f\rangle$, i.e., $G=G^T$. As for the complex inner-product, the adjoint of the operator $G$, denoted as $G^a$, is an operator such that $\left\langle f^*, G g\right\rangle=\left\langle g^*, G^a f\right\rangle^*$, and if the operator $G$ is self-adjoint or Hermitian, $\left\langle f^*, G g\right\rangle=\left\langle g^*, G f\right\rangle^*$, i.e., $G=G^a$\cite{chew1999waves}.

First, we derive the time-harmonic wave equation for the electric field. Suppose that the current source $\mathbf{J}$ radiates in an anisotropic, inhomogeneous medium, then the scattered time-harmonic fields denoted as $\mathbf{E}(\mathbf{r},t)=\mathbf{E}(\mathbf{r}) e^{i \omega t}$,  $\mathbf{H}(\mathbf{r},t)=\mathbf{H}(\mathbf{r}) e^{i \omega t}$ satisfy the Maxwell equations
	\begin{equation}
		\nabla \times \mathbf{E}(\mathbf{r})=-i \omega \mu_{0}\bar{\mu_{r}} \mathbf{H}(\mathbf{r}),
	\end{equation}
	\begin{equation}
		\nabla \times \mathbf{H}(\mathbf{r})=\mathbf{J}+i \omega \varepsilon_{0}\bar{\varepsilon_{r}} \mathbf{E}(\mathbf{r}),
	\end{equation}
	where the permittivity and permeability in free space are $\varepsilon_{0}$ and $\mu_{0}$, and the relative values of these two terms in the medium are $\bar{\varepsilon_{r}}$ and $\bar{\mu_{r}}$. $\nabla \times $ Eq. (1) and substitute the result into Eq. (2), we get
	\begin{equation}
		\nabla \times \bar{\mu_{r}}^{-1} \cdot \nabla \times \boldsymbol{\mathrm{E}(\mathbf{r})}-\omega^2 \bar{\varepsilon_{r}} \boldsymbol{\mathrm{E}(\mathbf{r})}=-i \omega \boldsymbol{\mathrm{J}}.
\end{equation}
If there are two groups of sources $\mathbf{J}_1$ and $\mathbf{J}_2$ produce the fields $\boldsymbol{\mathrm{E}}_1$ and $\boldsymbol{\mathrm{E}}_2$, then the vector wave equations are shown as follows, respectively
\begin{equation}
\nabla \times \bar{\mu}^{-1} \cdot \nabla \times \boldsymbol{\mathrm{E}}_1-\omega^2 \bar{\varepsilon} \boldsymbol{\mathrm{E}}_1=-i \omega \boldsymbol{\mathrm{J}}_1,
\end{equation}
\begin{equation}
\nabla \times \bar{\mu}^{-1} \cdot \nabla \times \boldsymbol{\mathrm{E}}_2-\omega^2 \bar{\varepsilon} \boldsymbol{\mathrm{E}}_2=-i \omega \boldsymbol{\mathrm{J}}_2.
\end{equation}
Taking the inner product of the Eq. (4) by $\boldsymbol{\mathrm{E}}_2$ and the Eq. (5) by $\boldsymbol{\mathrm{E}}_1$, we obtain
\begin{equation}
-i \omega\left\langle\boldsymbol{\mathrm{E}}_2, \boldsymbol{\mathrm{J}}_1\right\rangle=\left\langle\boldsymbol{\mathrm{E}}_2, \nabla \times \bar{\mu}^{-1} \cdot \nabla \times \boldsymbol{\mathrm{E}}_1\right\rangle-\omega^2\left\langle\boldsymbol{\mathrm{E}}_2, \bar{\varepsilon} \boldsymbol{\mathrm{E}}_1\right\rangle,
\end{equation}
\begin{equation}
-i \omega\left\langle\boldsymbol{\mathrm{E}}_1, \boldsymbol{\mathrm{J}}_2\right\rangle=\left\langle\boldsymbol{\mathrm{E}}_1, \nabla \times \bar{\mu}^{-1} \cdot \nabla \times \boldsymbol{\mathrm{E}}_2\right\rangle-\omega^2\left\langle\boldsymbol{\mathrm{E}}_1, \bar{\varepsilon} \boldsymbol{\mathrm{E}}_2\right\rangle.
\end{equation}
When $\bar{\mu}$ and $\bar{\varepsilon}$ are symmetric tensors, it can be derived that "$\nabla \times \bar{\mu}^{-1} \cdot \nabla \times$" is also a symmetric operator, i.e., the right hand sides of Eqs. (6) and (7) are symmetric about $\boldsymbol{\mathrm{E}}_1$ and $\boldsymbol{\mathrm{E}}_2$. Hence, the reciprocal theorem holds $\left\langle\boldsymbol{\mathrm{E}}_2,\boldsymbol{\mathrm{J}}_1\right\rangle=\left\langle\boldsymbol{\mathrm{E}}_1,\boldsymbol{\mathrm{J}}_2\right\rangle$. Obviously, the symmetry properties of the operators (i.e., transpose symmetric) are the necessary conditions to the reciprocal relation of system. 
\subsection{The orthogonality relation of waveguide modes}
For the waveguide that is invariant along the $z$ direction, we consider an electromagnetic wave having angular frequency $\omega$ and propagating in the $z$ direction with propagation constant $\beta$, where the electric and magnetic fields can be expressed as follows: 
\begin{equation}
\mathbf{E}(\mathbf{r}, t)=\left[\mathbf{e}_{t}(x, y)+\mathbf{z} e_{z}(x, y)\right] \exp (i\omega t- i\beta z),
\end{equation}
\begin{equation}
\mathbf{H}(\mathbf{r}, t)=\left[\mathbf{h}_{t}(x, y)+\mathbf{z} h_{z}(x, y)\right] \exp (i \omega t-i \beta z).
\end{equation}\\
Substituting the Eqs. (8) and (9) into the Maxwell's equations, the equations of transverse electric and magnetic fields could be written as 
\begin{equation}
\nabla_{t}^{2} \boldsymbol{\mathrm{{e}}}_{t}+\nabla_{t}\left(\boldsymbol{\mathrm{{e}}}_{t} \cdot \frac{\nabla_{t} n^{2}}{n^{2}}\right)+k_{0}^{2} n^{2} \boldsymbol{\mathrm{{e}}}_{t}=\beta^{2} \boldsymbol{\mathrm{{e}}}_{t},
\end{equation}
\begin{equation}
\nabla_{t}^{2} \boldsymbol{\mathrm{{h}}}_{t}+\frac{1}{n^{2}} \nabla_{t} n^{2} \times\left(\nabla_{t} \times \boldsymbol{\mathrm{{h}}}_{t}\right)+k_{0}^{2} n^{2} \boldsymbol{\mathrm{{h}}}_{t}=\beta^{2} \boldsymbol{\mathrm{{h}}}_{t},
\end{equation}
where $n$ is the refractive index of the material, and $k_{0}$ is the wave number in free space, $\mathbf{e}_{t}=\mathbf{x}e_{x}+\mathbf{y}e_{y}$, $\mathbf{h}_{t}=\mathbf{x}h_{x}+\mathbf{y}h_{y}$. They could also be expressed in the matrix form of operators like:  
\begin{equation}
p_{x x} e_{x}+p_{x y} e_{y}=\beta^{2} e_{x},
\end{equation}
\begin{equation}
p_{y x} e_{x}+p_{y y} e_{y}=\beta^{2} e_{y},
\end{equation}
\begin{equation}
q_{x x} h_{x}+q_{x y} h_{y}=\beta^{2} h_{x},
\end{equation}
\begin{equation}
q_{y x} h_{x}+q_{y y} h_{y}=\beta^{2} h_{y}.
\end{equation}
The operators are expanded as follows:
\begin{equation}
p_{x x} e_{x}=\frac{\partial}{\partial x}\left(\frac{1}{n_{1}^{2}} \frac{\partial\left(n_{1}^{2} e_{x}\right)}{\partial x}\right)+\frac{\partial^{2} e_{x}}{\partial y^{2}}+k_{0}^{2} n_{1}^{2} e_{x},\: 
p_{x y} e_{y}=\frac{\partial}{\partial x}\left(\frac{1}{n_{1}^{2}} \frac{\partial n_{1}^{2}}{\partial y} e_{y}\right),
\end{equation}
\begin{equation}
p_{y y} e_{y}=\frac{\partial}{\partial y}\left(\frac{1}{n_{1}^{2}} \frac{\partial\left(n_{1}^{2} e_{y}\right)}{\partial y}\right)+\frac{\partial^{2} e_{y}}{\partial x^{2}}+k_{0}^{2} n_{1}^{2} e_{y},\:p_{y x} e_{x}=\frac{\partial}{\partial y}\left(\frac{1}{n_{1}^{2}} \frac{\partial n_{1}^{2}}{\partial x} e_{x}\right),
\end{equation}
\begin{equation}
q_{x x} h_{x}=\frac{\partial^{2}}{\partial x^{2}} h_{x}+n_{2}^{2} \frac{\partial}{\partial y}\left(\frac{1}{n_{2}^{2}} \frac{\partial}{\partial y} h_{x}\right)+k^{2} n_{2}^{2} h_{x},\:q_{x y} h_{y}=-n_{2}^{2} \frac{\partial}{\partial y}\left(\frac{1}{n_{2}^{2}} \frac{\partial}{\partial x} h_{y}\right)+\frac{\partial^{2}}{\partial y \partial x} h_{y},
\end{equation}
\begin{equation}
q_{y y} h_{y}=\frac{\partial^{2}}{\partial y^{2}} h_{y}+n_{2}{ }^{2} \frac{\partial}{\partial x}\left(\frac{1}{n_{2}{ }^{2}} \frac{\partial}{\partial x} h_{y}\right)+k_{0}^{2} n_{2}{ }^{2} h_{y}, \: q_{y x} h_{x}=-n_{2}^{2} \frac{\partial}{\partial x}\left(\frac{1}{n_{2}^{2}} \frac{\partial}{\partial y} h_{x}\right)+\frac{\partial^{2}}{\partial x \partial y} h_{x}.
\end{equation}\\
Considering the guided mode electric field under $n_{1}=n_{r}+i \kappa$ and the guided mode magnetic field under the same system $n_{2}=n_{r}+i \kappa$, $n_{r}$ is the real part of the refractive index, which describes the phase velocity of light in the material, and $\kappa$ is the extinction coefficient, which quantifies the loss of wave amplitude as it propagates through the medium due to absorption or scattering. According to the integration by parts, one obtains
\begin{equation}
\left\langle\varphi_{1}, p_{x x} \varphi_{2}\right\rangle=\left\langle p_{x x}^{T} \varphi_{1}, \varphi_{2}\right\rangle=\left\langle q_{y y} \varphi_{1}, \varphi_{2}\right\rangle,
\end{equation}
where $\varphi$ refers to the scalar electric and magnetic field in the expanded equations. Therefore, we could derive that $p_{x x}^{T}=q_{y y}$, $p_{y y}^{T}=q_{x x}$, $p_{x y}^{T}=-q_{x y}$, $p_{y x}^{T}=-q_{y x}$. Multiplying $\left\langle h_{y 2}\right.$, $\left\langle h_{x 2}\right.$, $\left\langle e_{y 1}\right.$, $\left\langle e_{x 1}\right.$ by Eq. (12), (13), (14), (15), respectively, and subtracting one from the other yields the following equations:
\begin{equation}
\left\langle e_{y 1}, p_{x y}^{T} h_{y 2}\right\rangle-\left\langle e_{x 1}, q_{y x} h_{x 2}\right\rangle=\left(\beta_{2}{ }^{2}-\beta_{1}^{2}\right)\left\langle e_{x 1}, h_{y 2}\right\rangle,
\end{equation}
\begin{equation}
\left\langle e_{x 1}, p_{y x}^{T} h_{x 2}\right\rangle-\left\langle e_{y 1}, q_{x y} h_{y 2}\right\rangle=\left(\beta_{2}{ }^{2}-\beta_{1}^{2}\right)\left\langle e_{y 1}, h_{x 2}\right\rangle.
\end{equation}
Subtracting Eq. (22) from Eq. (21), we obtain $\left(\beta_{2}{ }^{2}-\beta_{1}{ }^{2}\right)\left(\left\langle e_{x 1}, h_{y 2}\right\rangle-\left\langle e_{y 1}, h_{x 2}\right\rangle\right)=0$. Thus, for $\beta_{1}^{2} \neq \beta_{2}^{2}$, we obtain the orthogonality relation
\begin{equation}
 \iint_{A} \boldsymbol{\mathrm{e}}_{j} \times \boldsymbol{\mathrm{h}}_{k} \cdot \boldsymbol{\mathrm{z}} d A=2 C \delta_{j k},
\end{equation}
\\
where A represents the cross-section that includes the waveguide and its vicinity at $z$ = const, $C$ is a normalization parameter, and $j$ and $k$ refer to the different orders of modes in the same system. As for its conjugate system
 $n_{2}=n_{r}-i \kappa$, we can also get $p_{x x}^{a}=q_{y y}$, $p_{y y}^{a}=q_{x x}$, $p_{x y}^{a}=-q_{x y}$, $p_{y x}^{a}=-q_{y x}$. The equation for conjugate integration could be written as
 $\left(\beta_{2}{ }^{2}-\left(\beta_{1}{ }^{2}\right)^{*}\right)\left(\left\langle e_{x 1}^{*}, h_{y 2}^{}\right\rangle-\left\langle e_{y 1}^{*}, h_{x 2}^{}\right\rangle\right)=0.$ In this case, the following conjugate orthogonality relation is satisfied for $\left(\beta_{1}^{2}\right)^{*} \neq \beta_{2}^{2}$
 \begin{equation}
 \iint_{A} \boldsymbol{\mathrm{e}}_{j}^{*} \times \boldsymbol{\mathrm{h}}_{k} \cdot \boldsymbol{\mathrm{z}} d A=2 C \delta_{j k},
 \end{equation}\\
where $j$ and $k$ refer to the different orders of modes in the two conjugate systems. This relationship also implies the power orthogonality. The equation (23) indicates that the loss or gain break the conjugate power orthogonality\cite{Morozko}, not only for PT-symmetric but also for arbitrary non-Hermitian systems. Remarkably, the power orthogonality may lost in the broken phase of the PT-symmetric waveguides. We will discuss this issue later on.

\subsection{The coupled mode equations for the PT-symmetric waveguides}
\begin{figure}[H]
        \centering
        \includegraphics[width=0.5\linewidth]{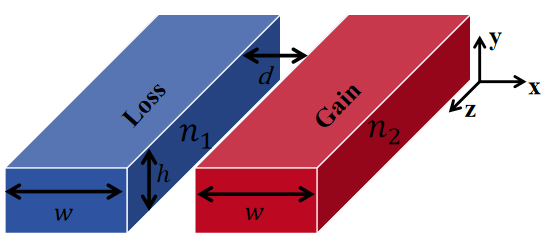}
        \caption{The schematic of PT-symmetric waveguides: Loss ($n_{1}=n_{r}-i \kappa$) and Gain ($n_{2}=n_{r}+i \kappa$) with the width $w$, height $h$, and the spacing distance $d$, embedded in the cladding medium.}
        \label{figure1}
\end{figure}

In this work, we focus on the mode power transmission properties of the PT-symmetric waveguides with the schematic shown in Fig. \ref{figure1}.

Next, we will derive the coupled mode equation based on the two orthogonality relations in the last section. The fields propagating in the $+z$ and $-z$ direction are denoted as $\phi$ = $\{\phi_{t}, \phi_{z}\}$ = $\{(e_{x}^{+}, e_{y}^{+}), e_{z}^{+}, (h_{x}^{+}, h_{y}^{+}), h_{z}^{+}\}$, $\psi$ = $\{\psi_{t}, \psi_{z}\}$ = $\{(e_{x}^{-}, e_{y}^{-}), e_{z}^{-}, (h_{x}^{-}, h_{y}^{-}), h_{z}^{-}\}$. Substituting the Eq. (8) and (9) of mode field expressions, the Maxwell equations could be written as:
\begin{equation}
\nabla_{t} \times \mathbf{e}_{z}^{+}-i \beta z \times \mathbf{e}_{t}^{+}=-i \omega \mu^{+} \mathbf{h}_{t}^{+},
\end{equation}
\begin{equation}
\nabla_{t} \times \mathbf{h}_{z}^{+}-i \beta z \times \mathbf{h}_{t}^{+}=i \omega \varepsilon^{+}\mathbf{e}_{t}^{+},
\end{equation}
\begin{equation}
\nabla_{t} \times \mathbf{e}_{z}^{-}+i \beta z \times \mathbf{e}_{t}^{-}=-i \omega  \mu^{-} \mathbf{h}_{t}^{-},
\end{equation}
\begin{equation}
\nabla_{t} \times \mathbf{h}_{z}^{-}+i \beta z \times \mathbf{h}_{t}^{-}=i \omega \varepsilon^{-}\mathbf{e}_{t}^{-},
\end{equation}
where $\varepsilon=\varepsilon_{0}\varepsilon_{r}$=$\{\varepsilon_{t}, \varepsilon_{z}\}$, $\mu=\mu_{0}\mu_{r}$=$\{\mu_{t}, \mu_{z}\}$. The matrix forms of the equations (25)(26), (27)(28) are expressed as $G^{+}\phi=0$ and $G^{-}\psi=0$, respectively. For the chiral symmetric materials, $\varepsilon_{t}^{-}$ = $\varepsilon_{t}^{+}$, $\varepsilon_{z}^{-}$ = $-\varepsilon_{z}^{+}$, while considering the conjugate case, $\varepsilon_{t}^{-}$ = $(\varepsilon_{t}^{+})^{*}$, $\varepsilon_{z}^{-}$ = $-(\varepsilon_{z}^{+})^{*}$, where $\ast$ indicates the operation of complex conjugation. Accordingly, it can be derived that $\mathbf{e}^{-}$, $\mathbf{h}^{-}$ are related to $\mathbf{e}^{+}$, $\mathbf{h}^{+}$ by $\mathbf{e}^{-}$ =  $\{e_{x}^{+}, e_{y}^{+}, -e_{z}^{+}\}$, $\mathbf{h}^{-}$ =  $\{-h_{x}^{+}, -h_{y}^{+}, h_{z}^{+}\}$ or $\mathbf{e}'^{-}$ =  $\{(e_{x}^{+})^{*}, (e_{y}^{+})^{*}, -(e_{z}^{+})^{*}\}$, $\mathbf{h}'^{-}$ =  $\{-(h_{x}^{+})^{*}, -(h_{y}^{+})^{*}, (h_{z}^{+})^{*}\}$. In order to establish the inner-product equation, the guided mode fields of the adjoint system are chosen as the test functions. Depending on the reciprocal and orthogonality relations discussed above, the different inner-product equations satisfy
$(\psi_{t},G^{+}\phi_{t})=(G^{-}\psi_{t},\phi_{t})$ or $(\psi_{t}^{'*},G^{+}\phi_{t})=((G_{a}^{-}\psi'_{t})^{*},\phi_{t})$, $G_{a}^{-}$ is the operator matrix in the conjugate case. Therefore we can expand them into the coupled mode equations. Firstly, we consider the weak coupling conditions and approximate the transverse total fields as the summation of the lossy guided mode fields $\{\mathbf{e}_{1t}^{+},\mathbf{h}_{1t}^{+}\}$ and the gain guided mode fields $\{\mathbf{e}_{2t}^{+},\mathbf{h}_{2t}^{+}\}$,  
$\boldsymbol{\mathrm{E}}_{t}^{+}\boldsymbol{(r)}=a_{1}^{+}(z) \boldsymbol{\mathrm{e}}_{1 t}^{+}+a_{2}^{+}(z) \boldsymbol{\mathrm{e}}_{2 t}^{-}, \boldsymbol{\mathrm{H}}_{t}^{+}\boldsymbol{(r)}=a_{1}^{+}(z) \boldsymbol{\mathrm{h}}_{1 t}^{+}+a_{2}^{+}(z) \boldsymbol{\mathrm{h}}_{2 t}^{+}$, the equations are as follows,
\begin{equation}
\nabla_{t} \times\left(\mathbf{e}_{t 1}^{+}+\mathbf{e}_{t 2}^{+}\right)+\frac{d a_{1}}{d z} z \times \mathbf{e}_{t 1}^{+}+\frac{d a_{2}}{d z} z \times \mathbf{e}_{t 2}^{+}=-i \omega \mu\left(\mathbf{h}_{t 1}^{+}+\mathbf{h}_{t 2}^{+}\right),
\end{equation}
\begin{equation}
\nabla_{t} \times\left(\mathbf{h}_{t 1}^{+}+\mathbf{h}_{t 2}^{+}\right)+\frac{d a_{1}}{d z} z \times \mathbf{h}_{t 1}^{+}+\frac{d a_{2}}{d z} z \times \mathbf{h}_{t 2}^{+}=i \omega \varepsilon\left(\mathbf{e}_{t 1}^{+}+\mathbf{e}_{t 2}^{+}\right).
\end{equation}
The fields of adjoint system within the single waveguide satisfy
\begin{equation}
\nabla_{t} \times \mathbf{e}_{z 1}^{-}+i \beta_{1} z \times \mathbf{e}_{t 1}^{-}=-i \omega \mu_{1}^{-} \mathbf{h}_{t 1}^{-},
\end{equation}
\begin{equation}
\nabla_{t} \times \mathbf{h}_{z 1}^{-}+i \beta_{1} z \times \mathbf{h}_{t 1}^{-}=i \omega \varepsilon_{1}^{-} \mathbf{e}_{t 1}^{-}.
\end{equation}
Following the non-conjugate inner-product equations and the conjugate inner-product equations, we can get
\begin{equation}
F=\mathbf{e}_{t 1}^{-}\cdot Eq. (30)-(\mathbf{e}_{t 1}^{+}+\mathbf{e}_{t 2}^{+})\cdot Eq. (32)+\mathbf{h}_{t 1}^{-}\cdot Eq. (29)-(\mathbf{h}_{t 1}^{+}+\mathbf{h}_{t 2}^{+})\cdot Eq. (31),
\end{equation}
\begin{equation}
F'=(\mathbf{e}_{t 1}^{'-})^{*}\cdot Eq. (30)-(\mathbf{e}_{t 1}^{+}+\mathbf{e}_{t 2}^{+})\cdot Eq'. (32)+(\mathbf{h}_{t 1}^{'-})^{*}\cdot Eq. (29)-(\mathbf{h}_{t 1}^{+}+\mathbf{h}_{t 2}^{+})\cdot Eq'. (31).
\end{equation}
Integrate over the cross-section, the coupled mode equation could be expressed uniformly as
\begin{equation}
\frac{d a_{1}^{+}}{d z}+c_{12}\left(\frac{d a_{2}^{+}}{d z}+i \beta_{2} a_{2}^{+}\right)=-i \beta_{1} a_{1}^{+}+i a_{1}^{+} k_{11}+i a_{2}^{+} k_{12}.
\end{equation}
Replacing the fields of Eqs. (31)(32) and taking the dot-product of Eqs. (29)(30) with $\mathbf{e}_{t 2}^{-}$, $\mathbf{h}_{t 2}^{-}$, the other equation could also be obtained
\begin{equation}
\frac{d a_{2}^{+}}{d z}+c_{21}\left(\frac{d a_{1}^{+}}{d z}+i \beta_{1} a_{1}^{+}\right)=-i \beta_{2} a_{2}^{+}+i a_{2}^{+} k_{22}+i a_{1}^{+} k_{21},
\end{equation}
where $k_{p q}=\frac{\omega \varepsilon_{0} \iint_{S_{d}}\left(\varepsilon_{r}-1\right) \boldsymbol{\mathrm{e}}_{p t}^{-} \cdot \boldsymbol{\mathrm{e}}_{q t}^{+} d x d y}{\iint_{-\infty}^{\infty}\left(\boldsymbol{\mathrm{h}}_{p t}^{+} \times \boldsymbol{\mathrm{e}}_{p t}^{-}+\boldsymbol{\mathrm{e}}_{p t}^{+} \times \boldsymbol{\mathrm{h}}_{p t}^{-}\right) d x d y}$, $c_{p q}=-\frac{\iint_{S_{d}}\left(\boldsymbol{\mathrm{h}}_{p t}^{-} \times \boldsymbol{\mathrm{e}}_{q t}^{+}+\boldsymbol{\mathrm{e}}_{p t}^{-} \times \boldsymbol{\mathrm{h}}_{q t}^{+}\right) d x d y}{\iint_{-\infty}^{\infty}\left(\boldsymbol{\mathrm{h}}_{p t}^{+} \times \boldsymbol{\mathrm{e}}_{p t}^{-}+\boldsymbol{\mathrm{e}}_{p t}^{+} \times \boldsymbol{\mathrm{h}}_{p t}^{-}\right) d x d y}$
 are from the $F$; $k'_{p q}=\frac{\omega \varepsilon_{0} \iint_{S_{d}}\left(\varepsilon_{r}-1\right) (\mathbf{e}^{'-}_{p t})^{*} \cdot \boldsymbol{\mathrm{e}}_{q t}^{+} d x d y}{\iint_{-\infty}^{\infty}\left(\boldsymbol{\mathrm{h}}_{p t}^{+} \times (\boldsymbol{\mathrm{e}}_{p t}^{'-})^{*}+\boldsymbol{\mathrm{e}}_{p t}^{+} \times (\boldsymbol{\mathrm{h}}_{p t}^{'-})^{*}\right) d x d y}$, $c'_{p q}=-\frac{\iint_{S_{d}}\left((\boldsymbol{\mathrm{h}}_{p t}^{'-})^{*} \times \boldsymbol{\mathrm{e}}_{q t}^{+}+(\boldsymbol{\mathrm{e}}_{p t}^{'-})^{*} \times \boldsymbol{\mathrm{h}}_{q t}^{+}\right) d x d y}{\iint_{-\infty}^{\infty}\left(\boldsymbol{\mathrm{h}}_{p t}^{+} \times (\boldsymbol{\mathrm{e}}_{p t}^{'-})^{*}+\boldsymbol{\mathrm{e}}_{p t}^{+} \times (\boldsymbol{\mathrm{h}}_{p t}^{'-})^{*}\right) d x d y}$ are from the $F'$; $p$ = 1 or 2, $q$ = 1 or 2; “d” denotes the perturbed region, the gain waveguide is regarded as the result of a perturbation to the cladding layer of the lossy waveguide, and vice versa. Thus for the Eqs. (35) and (36), the integration regions are the core section $S_{2}$, $S_{1}$ of the gain waveguide and the lossy waveguide, respectively. Notably, the value of $c_{p q}$ reflects the influence exerted by the rate of change and phase shift of mode $q$ on mode $p$, which represents a higher-order correction in the coupling dynamics. Compared to the direct coupling $k_{pq}$, the magnitude of this additional influence is negligible, and the actual calculated values of our examples confirm that. 
 
 For comparison, we derive the coupled mode equations based on the CCMT\cite{haus1987coupled}, the inner-product equation is $(\psi_{t}^{*},G^{+}\phi_{t})=((G^{-}\psi_{t})^{*},\phi_{t})$ or $(\psi'_{t},G^{+}\phi_{t})=(G_{a}^{-}\psi'_{t},\phi_{t})$. Similarly, we can get:\\
 \begin{equation}
\frac{d a_{1}^{+}}{d z}+c_{12}\left(\frac{d a_{2}^{+}}{d z}+i \beta_{2} a_{2}^{+}\right)=-i \beta_{1}^{*} a_{1}^{+}+i a_{1}^{+} k_{11}+i a_{2}^{+} k_{12},
\end{equation}
\begin{equation}
\frac{d a_{2}^{+}}{d z}+c_{21}\left(\frac{d a_{1}^{+}}{d z}+i \beta_{1} a_{1}^{+}\right)=-i \beta_{2}^{*} a_{2}^{+}+i a_{2}^{+} k_{22}+i a_{1}^{+} k_{21},
\end{equation}
where $k_{p q}=\frac{\omega \varepsilon_{0} \iint_{S_{d}}\left(\varepsilon_{r}-1\right) (\boldsymbol{\mathrm{e}}_{p t}^{-})^{*} \cdot \boldsymbol{\mathrm{e}}_{q t}^{+} d x d y}{\iint_{-\infty}^{\infty}\left(\boldsymbol{\mathrm{h}}_{p t}^{+} \times (\boldsymbol{\mathrm{e}}_{p t}^{-})^{*}+\boldsymbol{\mathrm{e}}_{p t}^{+} \times (\boldsymbol{\mathrm{h}}_{p t}^{-})^{*}\right) d x d y}$, $c_{p q}=-\frac{\iint_{S_{d}}\left((\boldsymbol{\mathrm{h}}_{p t}^{-})^{*} \times \boldsymbol{\mathrm{e}}_{q t}^{+}+(\boldsymbol{\mathrm{e}}_{p t}^{-})^{*} \times \boldsymbol{\mathrm{h}}_{q t}^{+}\right) d x d y}{\iint_{-\infty}^{\infty}\left(\boldsymbol{\mathrm{h}}_{p t}^{+} \times (\boldsymbol{\mathrm{e}}_{p t}^{-})^{*}+\boldsymbol{\mathrm{e}}_{p t}^{+} \times (\boldsymbol{\mathrm{h}}_{p t}^{-})^{*}\right) d x d y}$ are from the $F$, $F'$ follows a similar procedure. The results in the next section will show that the CCMT fails to capture the mode power transmission properties of the PT-symmetric waveguide.

The coupled mode equations provide the analytic references for the mode power transmissions of the PT-symmetric waveguide and it can be further applied to the other directional couplers. Regarding to the numerical or measurement results, in order to separate the contribution of the guided mode from any other radiative scattering, the particular guided mode field $\boldsymbol{\mathrm{E}}_{m}$ and $\boldsymbol{\mathrm{H}}_{m}$ are computed first. Then the power $P_{m}$ carried in $\boldsymbol{\mathrm{E}}$ and $\boldsymbol{\mathrm{H}}$ by the mode from the conventional method is given by
\begin{equation}
P_{m}=\left|\frac{1}{4} \iint_{A}\left(\boldsymbol{\mathrm{E}}_{m}^{*} \times \boldsymbol{\mathrm{H}}+\boldsymbol{\mathrm{E}} \times \boldsymbol{\mathrm{H}}_{m}^{*}\right) \cdot d \boldsymbol{A}\right|^{2},
\end{equation}
based on the well-known orthogonality properties of the modes at a fixed frequency\cite{marcuse2013theory19,skorobogatiy2009fundamentals20,johnson2002adiabatic21}.
This study concentrates on the calculation of power transmission of complex mode field $\boldsymbol{\mathrm{E}}_{lm}$ and $\boldsymbol{\mathrm{E}}_{gm}$ in the lossy/gain waveguides, where the power of lossy mode $P_{lm}$ is calculated by 
\begin{equation}
P_{lm}=\left|\frac{1}{4} \iint_{A}\left(\boldsymbol{\mathrm{E}}_{lm} \times \boldsymbol{\mathrm{H}}+\boldsymbol{\mathrm{E}} \times \boldsymbol{\mathrm{H}}_{lm}\right) \cdot d \boldsymbol{A}\right|^{2},
\end{equation}
based on the non-conjugate orthogonality relation as discussed above. This equation, corresponding to the conjugate orthogonality relation, can also be written in the form of the conjugate mode as follows:
\begin{equation}
P_{lm}=\left|\frac{1}{4} \iint_{A}\left(\boldsymbol{\mathrm{E}}_{g m}^{*} \times \boldsymbol{\mathrm{H}}+\boldsymbol{\mathrm{E}} \times \boldsymbol{\mathrm{H}}_{g m}^{*}\right) \cdot d \boldsymbol{A}\right|^{2}.
\end{equation}

In combination with the above two differential Eqs. (35) and (36), one variable can be eliminated such that
\begin{equation}
\frac{d^{2} a_{1}}{d z^{2}}+i\left(\gamma_{1}+\gamma_{2}\right) \frac{d a_{1}}{d z}+\left(k_{12} k_{21}-\gamma_{1} \gamma_{2}\right) a_{1}=0 \text {, }
\end{equation}
where $\gamma_{1}=k_{11}+\beta_{1}$, $\gamma_{2}=k_{22}+\beta_{2}$. The eigenvalues of the exponential term are
\begin{equation}
\lambda_{ \pm}=\frac{1}{2} i\left(\gamma_{1}+\gamma_{2}\right) \mp \frac{i}{2} \sqrt{\left(\gamma_{1}-\gamma_{2}\right)^{2}+4 k_{12} k_{21}} .
\end{equation}
When varying two of the available parameters, the two eigenvalues can be forced to  coalesce at a specific point called the “exceptional point” (EP)\cite{ozdemir2019parity,chatzidimitriou2018optical} where $\left(\gamma_{1}-\gamma_{2}\right)^{2}+4 k_{12} k_{21}=0$. The system exhibits two phases: PT-symmetry phase and PT-broken phase. The system undergoes a transition from a completely real spectrum into a complex spectrum, which is known as PT-symmetry breaking\cite{S.Klaiman,P.Chen,Lu2023guided}. Moreover, the breaking of PT symmetry is not only limited to gain-loss systems but can also occur in other types of systems\cite{zhang2023gain,chatzidimitriou2021breaking}. Next, we will discuss the mode power transmission and phase transition properties of the PT-symmetric waveguides in detail, as well as the conserved quantity for the Hamiltonian invariant of the system.

\section{Results and discussions}
\subsection{The eigenmodes of PT-symmetric waveguides}
\begin{figure}[ht]
    \centering
    \includegraphics[width=1\linewidth,height=0.25\textheight]{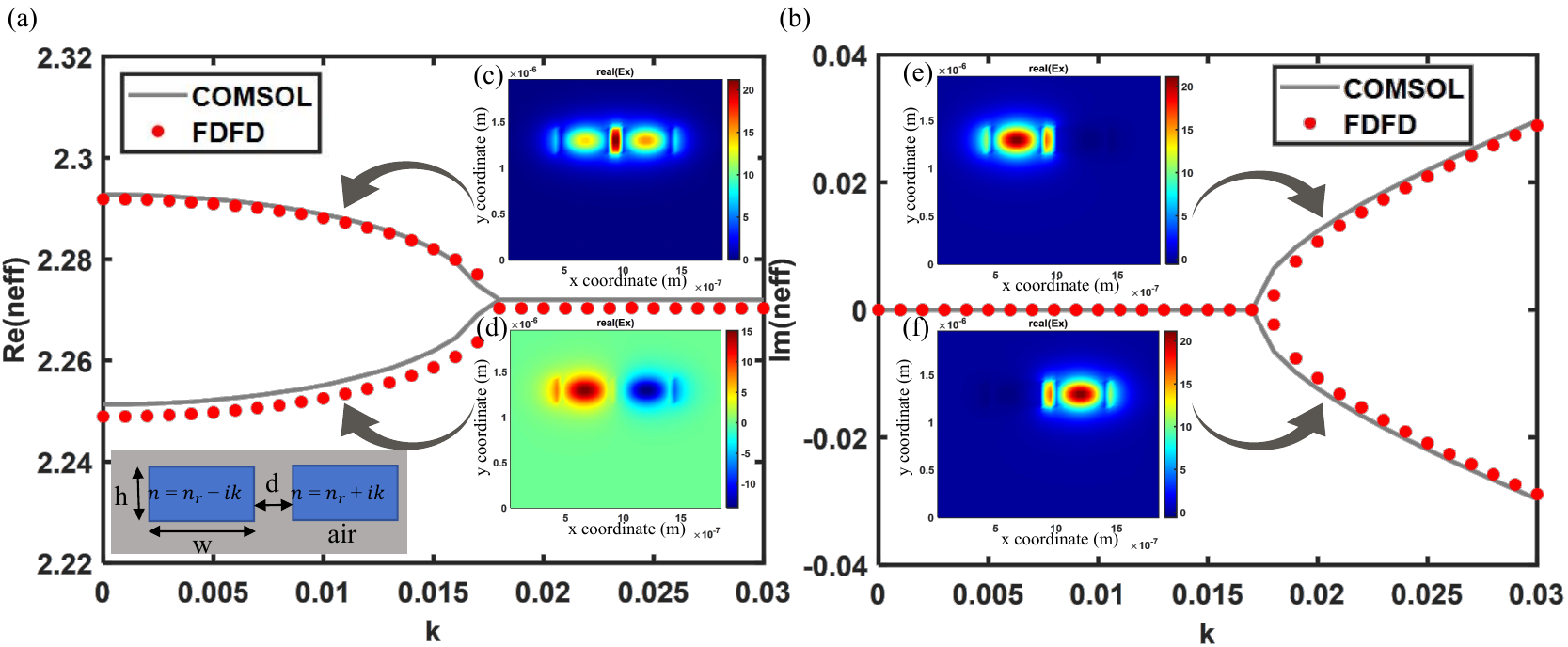}
    \caption{Real part (a) and imaginary part (b) of the effective mode index ($n_{eff}$) versus $k$ calculated by the FDFD method (red dots) and COMSOL (Gray solid lines). (c)(d) Re$\{$$E_{x}$$\}$ for “symmetric” and “antisymmetric” supermodes at $k=0.008$; (e)(f) Re$\{$$E_{x}$$\}$ for
    modes with loss and gain at $k=0.05$,  respectively.}
    \label{neff}
\end{figure}

\begin{figure}[h]
    \centering
    \includegraphics[width=0.95\textwidth,height=0.25\textheight]{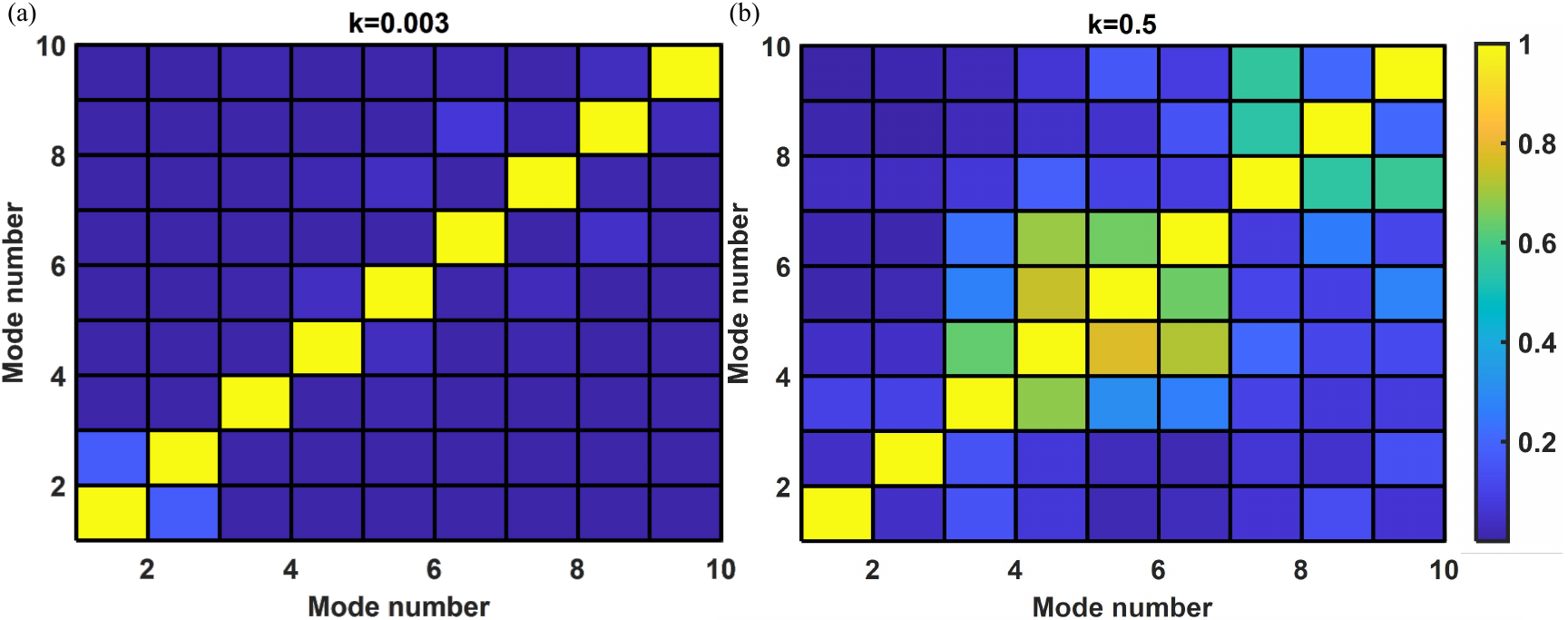}
    \caption{The diagram of mode power orthogonality under  (a) k = 0.003, (b) k = 0.5.}
    \label{orthogonality}
\end{figure}

In this part, we use the FDFD method to study the dispersion relations of the coupled waveguides with balanced loss and gain\cite{Ctyro}. In comparison to the time-domain methods encountering divergence issues, this method demonstrates strong stability for solving complex modes and the ability to handle complex boundary conditions in non-Hermitian systems. Additionally, the FDFD method enables the direct calculation of the transmission parameters, facilitating the analysis of mode interactions and the coupling strengths, which is well-suited for solving the multimode problems and evaluating the contrast transmission behaviors before and after phase transition for the PT-symmetric waveguide. The following material and geometrical parameters corresponding to the general silicon-based optical waveguide devices are chosen: $n_{r}=3.48$, $w=0.45$ $\mu m$, $h=0.225$ $\mu m$, $d=0.09$ $\mu m$; the working wavelength is $\lambda=1.55$ $\mu m$ and $k$ is the independent variable. The schematic is shown in the inset of Fig. \ref{neff} (a). It is known that a pair of even and odd supermodes emerge in this instance. Numerically calculated real and imaginary parts of the effective refractive index ($n_{eff}$) of both two eigenmodes as a function of $k$ are shown in Figs. \ref{neff} (a), (b), respectively. They match results from COMSOL remarkably well, which demonstrates the accuracy of the algorithm and the results. As discussed in Section 2.3, the EP is a single point, and the corresponding $k$ value of this structure at this point is around 0.018. Below this critical value, the guided modes propagate with real $n_{eff}$, while beyond the EP, the imaginary part of the $n_{eff}$ splits into two branches. Transversal distributions of the eigenmode fields for the different values of $k$ below and above the EP, are plotted in Figs. \ref{neff} (c)(d), (e)(f). It is obvious that phase transition occurs in the fields at the EP. 
The mode confinement and the propagation constant exhibit a substantial difference before and after the phase transition of the PT-symmetric system. The whole system is under the PT-symmetric and broken phase states, respectively. For exploring the issue of the mode power transmission, it is worthwhile to analyze the mode orthogonality of the system. The power flows between different modes are calculated by the Eq. (23) under the different $k$ = 0.003, and 0.5 in the Fig. \ref{orthogonality}. It can be observed that the mode powers exhibit great orthogonality in the PT-symmetric phase. When the system enters the PT-broken phase, this orthogonality is lost, and the inner product of mode power becomes non-zero. Thus, it can be inferred that the different mode power flows remain independent, which makes the gain and loss of the system power balanced. If the mode power coupling occurs, the system power may exhibit unstable behaviors. In the section that follows, it will be argued that the mode power transmission properties of PT-symmetric waveguides under different system states corresponding to various $k$ values.
\begin{figure}[h]
    \centering
    \includegraphics[width=1\linewidth]{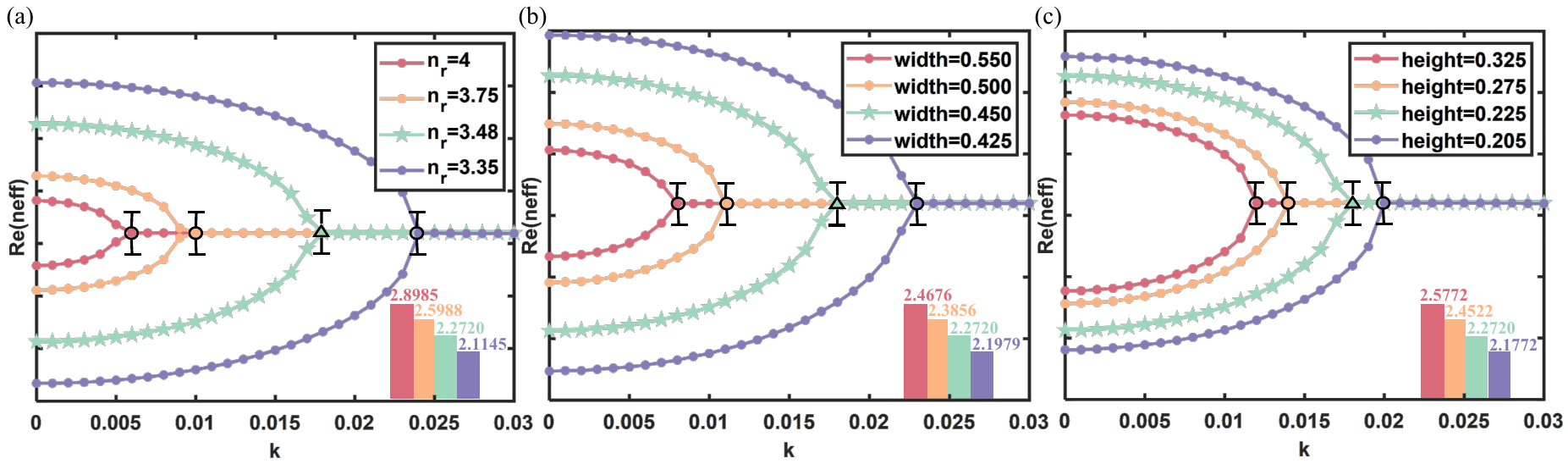}
    \caption{Real part of the $n_{eff}$ aligned to a common line with different (a) $n_{r}$, (b) width, (c) height of the waveguide. The actual values of which are displayed in bar charts.}
    \label{nr}
\end{figure}

In order to study the effects of different structural parameters on the EP thoroughly, we calculate the $n_{eff}$ of eigenmodes under the different material parameters, $n_{r}$ = 4.00, 3.75, 3.48, 3.35; $w $ = 0.55 $\mu m$, 0.5 $\mu m$, 0.45 $\mu m$, 0.425 $\mu m$; and $h$ = 0.325 $\mu m$, 0.275 $\mu m$, 0.225 $\mu m$, 0.205 $\mu m$. The real parts of the calculated $n_{eff}$ are depicted in Fig. \ref{nr}. The real parts of $n_{eff}$ at the EP have been aligned to a common line for easier comparison, the actual values of which are displayed in bar charts. The green curve is the result for the original waveguide parameters: $n_{r} = 3.48$, $w$ = 0.45 $\mu m$, and $h$ = 0.225 $\mu m$. The results show that when the $n_{r}$, width, and height of the waveguide increases individually, the real part of the $n_{eff}$ increases and the $k$ value of the structure decreases at the EP. If we need smaller $k$ value of the material parameters to achieve the phase transition, the value of the $n_{r}$, width, and height of the waveguide could be designed to be larger. This finding could be useful for the engineering design of PT-symmetric waveguides. Nevertheless, there exist several structural manufacturing challenges that need to be addressed. Firstly, gain and loss are often influenced by external factors, such as temperature fluctuations, fabrication tolerances, and material imperfections, making it difficult to achieve the necessary precision in control. Secondly, high gain can introduce noise or lead to material degradation over time, which impacts the reliability and longevity of devices. Thirdly, achieving the required geometrical and refractive index precision during fabrication is challenging, especially as nanofabrication techniques are subject to imperfections and variability. Finally, if the system operates at an EP, even small perturbations can lead to large changes in the output, which can be beneficial for sensing but problematic for switching devices that require robustness.\\
 
\subsection{The extraction of transmission parameters}
\begin{figure}[t]
    \centering
    \includegraphics[width=0.6\textwidth,height=0.25\textheight]{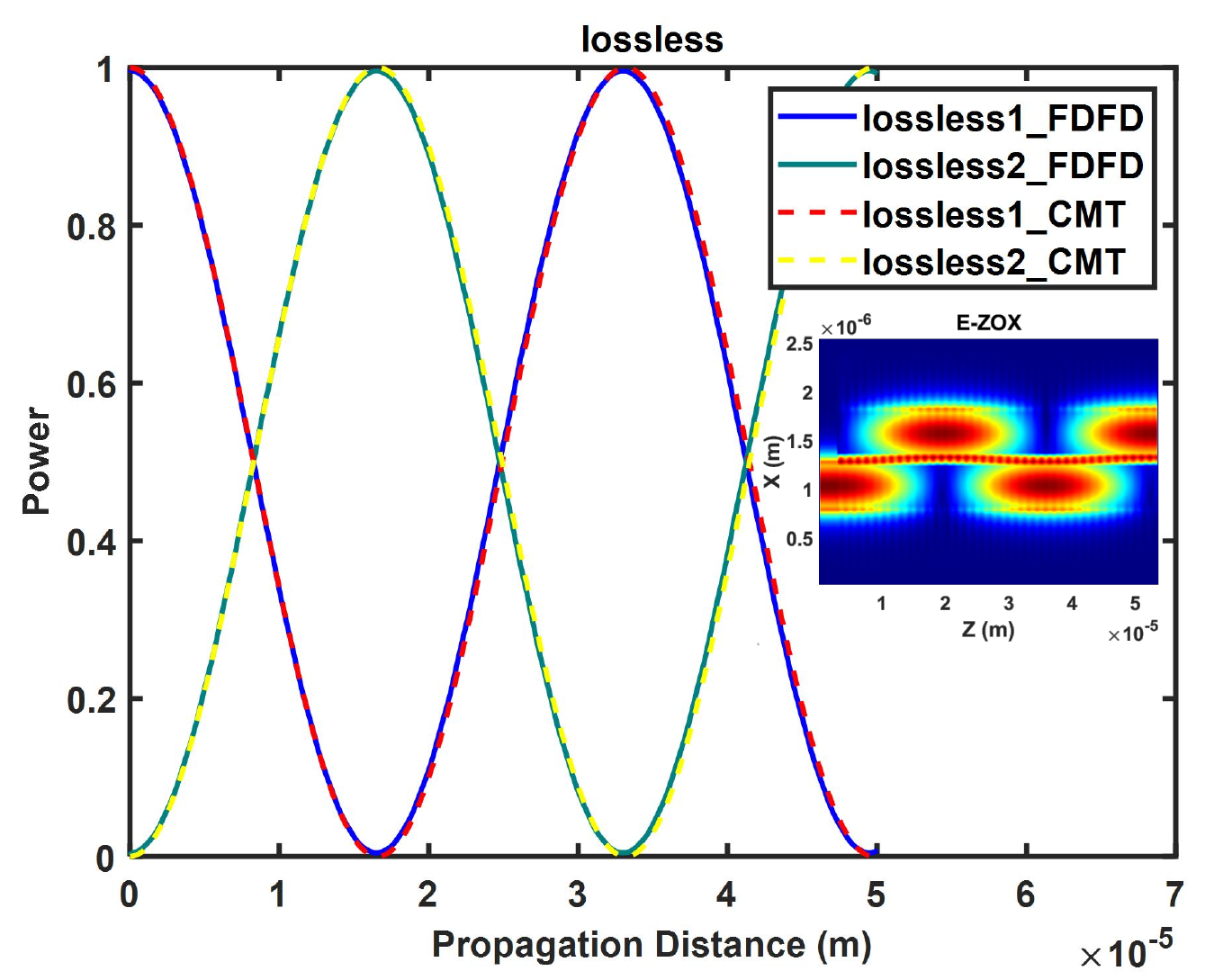}
    \caption{Power exchange of lossless waveguides as a function of the propagation distance, where the solid curves are obtained from the FDFD method while the dashed curves from the CMT. The inset shows the distribution of transmitted electric field.}
    \label{lossless}
\end{figure}

\begin{figure}[t]
    \centering
    \includegraphics[width=1\linewidth]{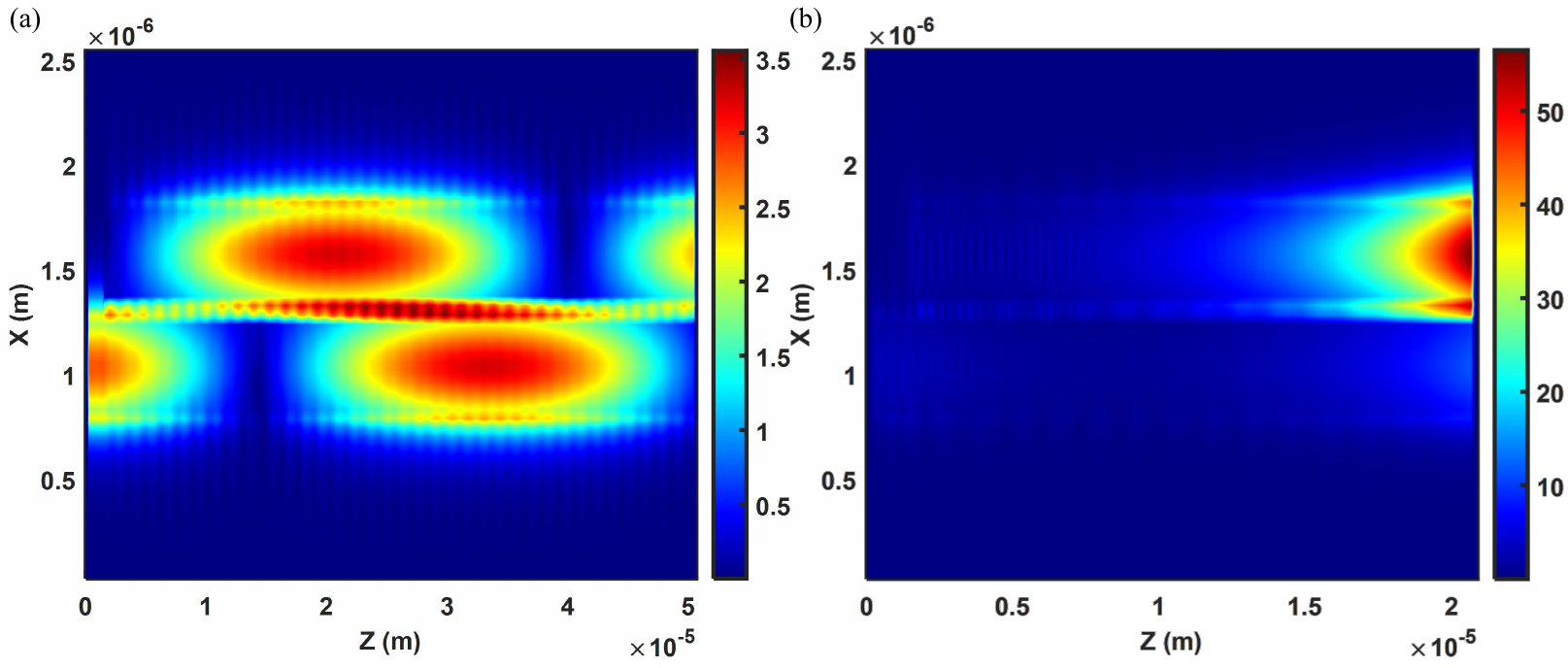}
    \caption{Electric field distribution $\left|\boldsymbol{\mathrm{E}}_{\text {total}}(x, z)\right|$ along the propagation direction of the PT-symmetric waveguides at (a) $k=0.01$ and (b) $k=0.05$.}
    \label{field_plot}
\end{figure}

\begin{figure}[t]
    \centering
    \includegraphics[width=1\linewidth]{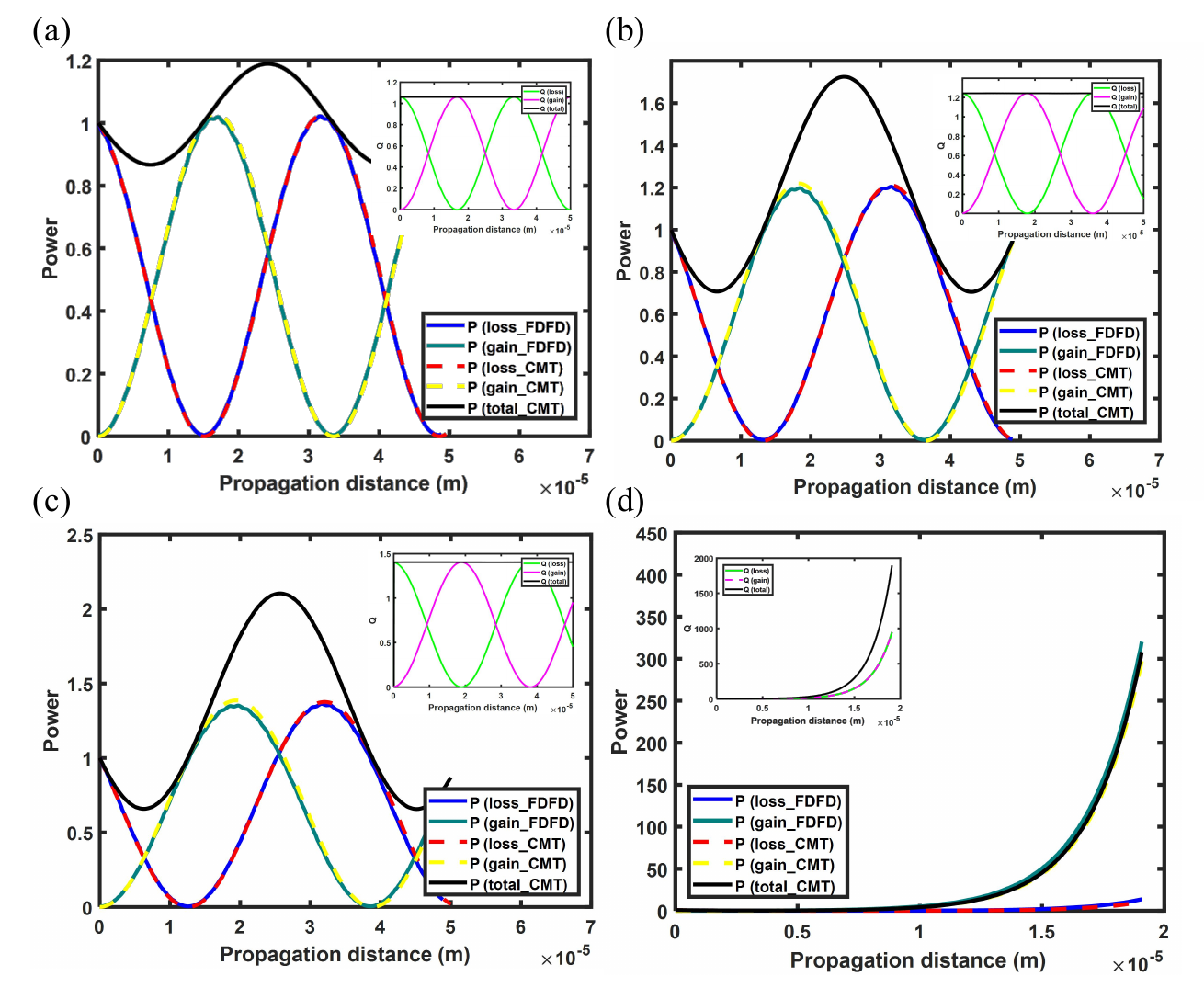}
   \caption{The lossy and gain mode power transmission at different $k$ (a) 0.003; (b) 0.008; (c) 0.01; (d) 0.05. The main panels show the power in the two waveguides as a function of the propagation distance calculated by the CMT and the FDFD method. The results presented in the insets correspond to the conserved quantity calculated by the composite modes $\phi_{lm}$ and $\phi_{gm}$.}
    \label{power_q}
\end{figure}

\begin{figure}
    \centering
    \includegraphics[width=1\linewidth]{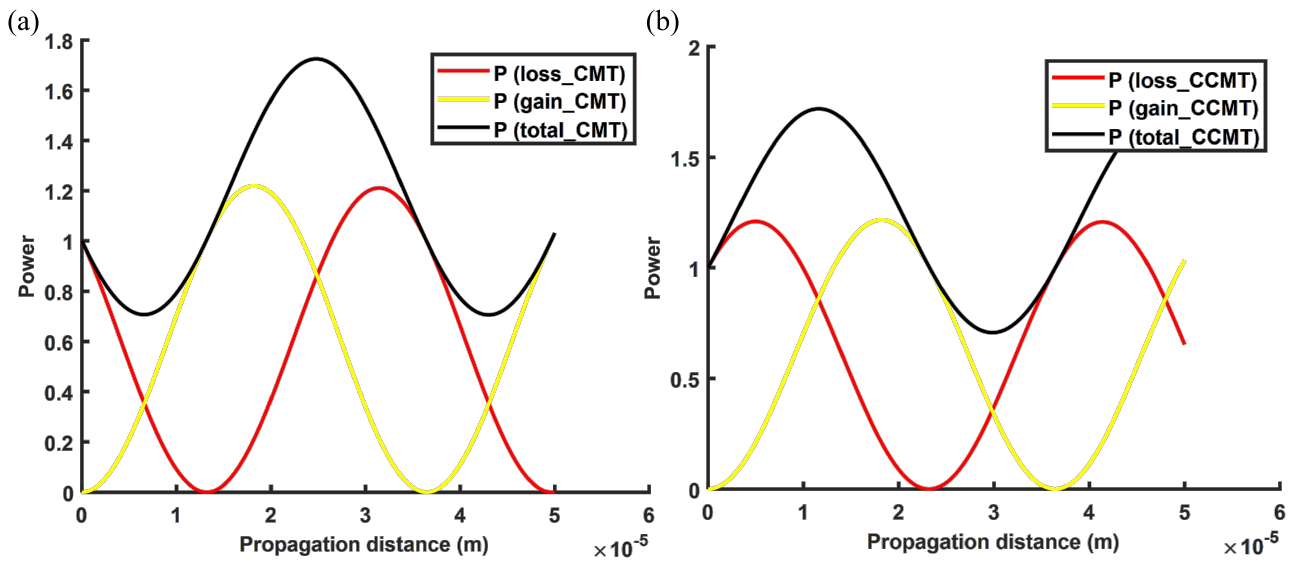}
    \caption{The lossy and gain mode power transmission at $k$ = 0.008 using the (a) CMT and (b) CCMT.}
    \label{ccmt}
\end{figure}

\begin{figure}
    \centering
    \includegraphics[width=1\linewidth]{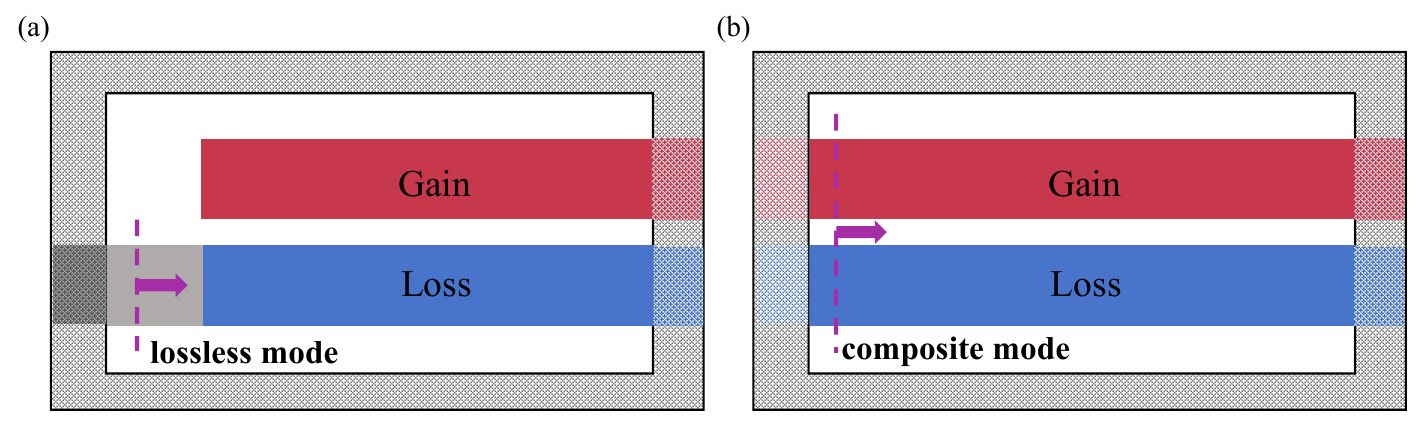}
    \caption{The schematic diagram of excitation by the different mode sources. (a) lossless mode source input from the lossless medium and (b) composite localized mode source input from the left side of the whole structure.}
    \label{pml}
\end{figure}

Based on the aforementioned CMT, we first calculate the coupling coefficients and the power transmission of the lossless coupled waveguides. The FDFD numerical simulation results show excellent agreement with the theoretical results as shown in Fig. \ref{lossless}, where the relative L2-norm error is $2\%$. The effective indices of the symmetric and antisymmetric supermodes are calculated to be $N_{s}=2.2863$ and $N_{a}=2.2404$. Hence, the coupling length can be calculated\cite{Huang} using the equation $L_{c}=\frac{\lambda}{2\left(N_{s}-N_{a}\right)}$, where $\lambda$ is the wavelength. The result is equal to 16.885 $\mu m$. For the PT-symmetric waveguides, the electric field distributions $\left|\boldsymbol{\mathrm{E}}_{\text {total }}(x, z)\right|$ by the FDFD is plotted in Fig. \ref{field_plot}(a) for $k=0.01$ (below the critical value) and in Fig. \ref{field_plot}(b) for $k=0.05$ (above the critical value). 
The mode power transmissions at different $k$ values are shown in Fig. \ref{power_q}, the numerical result for the extraction of transmission parameters is performed by the cross product of the original guided mode field and the total field by the Eq. (40) or (41), coincides with the theoretical results. It is evident that the waveguide coupler exhibits significantly different behaviors when the gain/loss coefficient is below and above the critical value in Fig. \ref{field_plot}. Below the critical value, light exchanges periodically between the two waveguides, but in an asymmetric manner. Above the critical value, the behavior of coupler changes significantly, where most of the power concentrates in the gain waveguide, fitting a typical feature of PT-broken phase. Accordingly, the power transmission curves show the asymmetric oscillations and the monotonic increasing properties respectively in Fig. \ref{power_q}. It also shows that the oscillations of the curve become more intense with the increase of k below the critical value. We calculate the relative L2-norm errors to be 4.16$\%$, 4.81$\%$, and 6.39$\%$ with the increase of $k$ value. The reason for the larger errors might be the instability and evolution characteristics of guided mode fields approaching the EP. In summary, in the PT-symmetric phase, the eigenmodes have real eigenvalues, and the transmission modal fields are generally well-confined, as the gain and loss profiles are carefully balanced. And the mode transmission parameters are typically symmetric. In the PT-broken phase, non-reciprocal scattering occurs. The gain in one direction and loss in the other causes unidirectional transmission or reflection. The group velocity dispersion can become highly anomalous.
In order to validate the accuracy of the theoretical method, we also compare the power transmission results by the CCMT as analyzed in Section 2.3. The Fig. \ref{ccmt} exhibits that the CCMT does not apply to the non-Hermitian problem of PT-symmetric waveguide. The propagation constant of the complex conjugate in the Eqs. (37) and (38) introduces error of the phase shift. For the real propagation constant of the lossless waveguide, the complex conjugate will not produce any additional effects and the error will be eliminated. Therefore, it can be concluded that the CCMT works fine for the lossless system, but fails to capture the major power features of the PT-symmetric waveguide in our case. The CMT we used has successfully obtained the power transmission properties.\\
Furthermore, we can observe that the sum of the two mode powers exhibits oscillation characteristics during transmission in Fig. \ref{power_q}. It corresponds to the actual power of the system defined by $P=\iint_{s}(\mathbf{E}(x, y) \times \mathbf{H}^{*}(x, y)+\mathbf{E}^{*}(x, y) \times \mathbf{H}(x, y)) dxdy$, where the integration region $S$ refers to the cross section of the whole structure including the two waveguides and the surrounding air. It is to be noted that the symmetry transformation is related to the Hamiltonian of the system. For the PT-symmetric system, although the Hamiltonian of the system may be non-Hermitian, it possesses symmetry referring to invariance under parity (P) and time reversal (T) operations. Hence, there must exist a conserved quantity taking forms analogous to the power flow, which is a constant of motion independent of distance $z$ when the PT-symmetry is unbroken. It is intrinsically linked to the Hamiltonian invariant of the system. In this work, we express the quantity as $Q=\iint_{s}(\mathbf{E}(x, y) \times \mathbf{H}^{*}(-x, y)+\mathbf{E}^{*}(-x, y) \times \mathbf{H}(x, y))dxdy$ \cite{S.Klaiman}.
In order to explore the individual mode power transmission properties in the conserved quantity, we refer to the calculation of the test localized transverse guided mode fields $\mathbf{E}_{lm}$ and $\mathbf{E}_{gm}$ in the lossless system. They could be equivalently regarded as the sum and difference of the two system supermodes. Thus, we perform similar operations on the supermodes of the PT-symmetric system, i.e., $\mathbf{E}_{lm}^{'}=\mathbf{E}_{mode1}+\mathbf{E}_{mode2}$, $\mathbf{E}_{gm}^{'}=\mathbf{E}_{mode1}-\mathbf{E}_{mode2}$, where the $\mathbf{E}_{mode}$ refers to the transverse guided mode field of the entire structure. Then the power of "lossy mode" in the conserved quantity could be obtained by $Q_{lm}=\iint_{s}(\mathbf{E}_{lm}^{'}(x, y) \times \mathbf{H}^{*}(-x, y) +\mathbf{E}^{*}(-x, y) \times \mathbf{H}_{lm}^{'}(x,y))dxdy$. In this case, we choose $\mathbf{E}_{lm}^{'}$ as the excitation source for the entire system, which differs from the previous case that uses the guided mode field of the single lossless waveguide as the excitation source. The schematic diagram is shown in Fig. \ref{pml}. The gray shaded area denotes the stretched-coordinate perfectly matched layer absorbing boundary conditions at all boundaries of the simulation domain\cite{jin2015finite,shin2012choice,chew1997complex}. 

In the numerics, $Q_{lm}$ and $Q_{gm}$ are calculated and the results are shown in the insets of Fig. \ref{power_q}. It can be observed that in the PT-symmetric system and below the critical value, the $Q_{lm}$ and $Q_{gm}$ curves show symmetric transmission behaviors. The distinction between the curves plotted by $Q_{lm}$, $Q_{gm}$ and $P_{lm}$, $P_{gm}$ is characterized by the phase differences. And the conserved Q value is equal to the average of the integrated actual power of the original modes. It indicates that the introduction of the permittivity imaginary part with opposite signs causes the oscillation, and the equilibrium value of the oscillation is associated with the Hamiltonian invariant of the system. In addition, the coupling lengths could be calculated as 1.703 $\mu m$, 1.85 $\mu m$, and 1.97 $\mu m$ by the formula $L_{c}=\frac{\lambda}{2\left(N_{s}-N_{a}\right)}$, which are consistent with the periodic lengths of the two symmetric curves of $Q_{lm}$ and $Q_{gm}$ for different $k$ in the insets. For larger $k$ above the critical value, the original conservation relation is broken, but the integrated quantities of $Q_{lm}$, $Q_{gm}$ maintain an equal increasing trend. Besides, we study the symmetric lossy and gain dual waveguides. Differing from the PT-symmetric system, the quantities of $Q_{lm}$ and $Q_{gm}$ equal to the two actual mode powers in the individual waveguides with no phase difference, both of which agree well with the theoretical solutions of CMT as shown in Fig. 10. The sum of the power of the two modes exhibits the decreasing trend and increasing trend respectively, which are distinct from the oscillation characteristics and the conserved quantity of the PT-symmetric system. The results show that the conserved quantity exists in the Hermitian system, as well as in the system of Hamiltonian invariant, but is not present in the normal lossy system.
\begin{figure}
    \centering
    \includegraphics[width=1\linewidth]{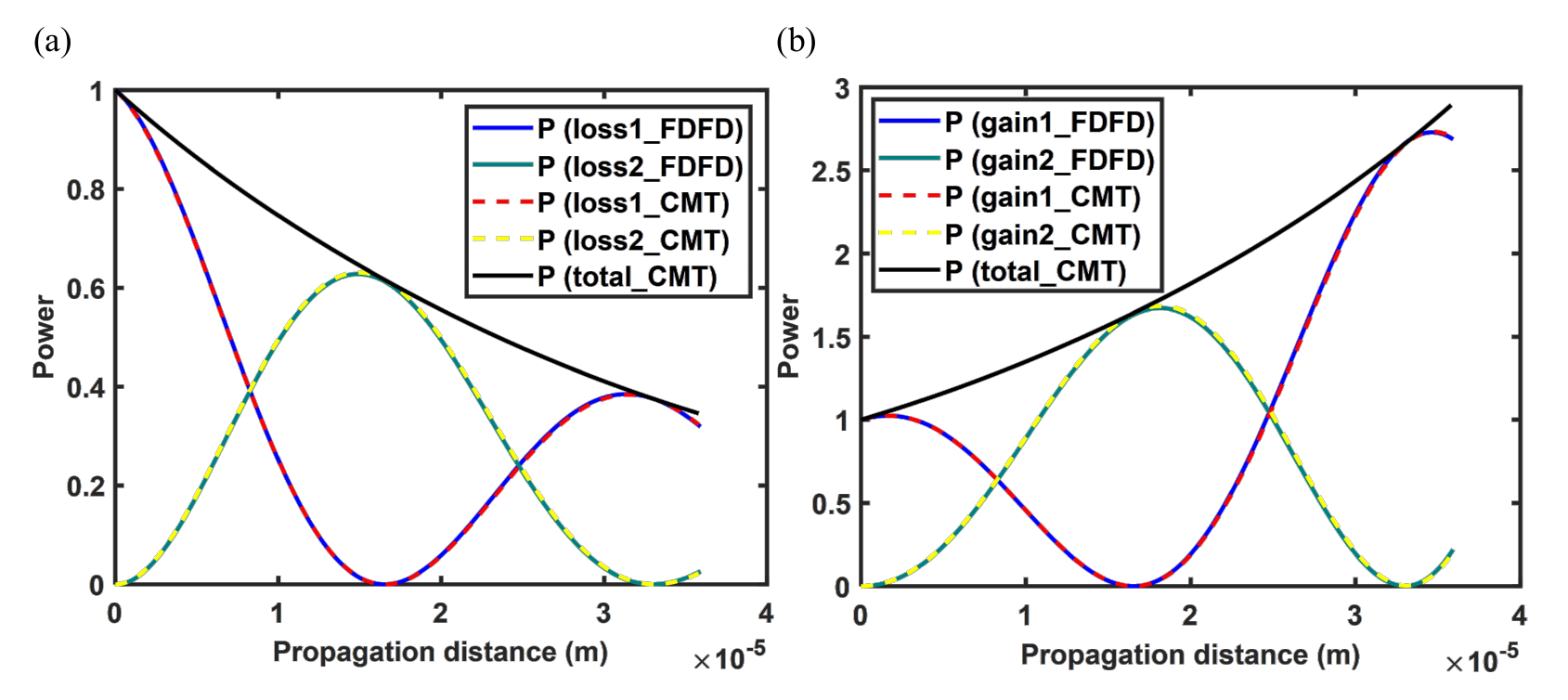}
     \caption{The mode powers of the (a) lossy dual waveguides and (b) gain dual waveguides at k=0.003.}
    \label{loss_gain}
\end{figure}

\section{Conclusions}
Prior work has documented the generalized coupled mode formalism in reciprocal waveguides with gain, loss and studied the dispersion of the PT-symmetric waveguide. However, these studies have not focused on the mode power transmission or the system power properties of such non-Hermitian structure. In this work, we 
derive the different orthogonality relations of complex modes and construct the coupled mode equations between the original and adjoint system to extract the transmission parameters. The orthogonality and conserved properties of system power under the phase transition are also analyzed. We find that the power transmission results from the coupled mode equations show excellent agreement with numerical simulations, while the results from CCMT fail to capture the features. Moreover, the mode power transmission within the individual waveguide exhibits periodic oscillations and the system power displays orthogonality, both phenomena are disrupted as the EP is approached. The conserved quantity Q is observed in the PT-symmetric phase, which is consistent with the average sum of the actual mode power. These findings extend the employment of the CMT in the non-Hermitian structure and demonstrate the extraordinary power properties of the 
PT-symmetric waveguide. In addition, the lossy and gain mode power of the conserved quantity are extracted by the combination of the supermodes, which provide a new perspective for the analysis of conserved quantity. The symmetric lossy or gain dual waveguides are also calculated to exhibit the decreasing and increasing trend of system power differing from the PT-symmetric waveguide, which show the effect of the extinction coefficients on the system power. Our work will be helpful for the direct analysis of the mode power transmission in the PT-symmetric system. Furthermore, the specific power properties of the non-Hermitian system will pave a new way for designing the various optical waveguide structures or light switching applications. The future work should include the power evaluation of multimode or hybrid modes for more complicated structures.

\begin{backmatter}
\bmsection{Funding} 
National Key Research and Development Program of China (2021YFB2800302).

\bmsection{Disclosures} The authors declare that there are no conflicts of interest related to this article.

\bmsection{Data availability} Data underlying the results presented in this paper are not publicly available at this time but may
 be obtained from the authors upon reasonable request.

\end{backmatter}

%%%%%%%%%% If using BibTeX:
\bibliography{main}

%%%%%%%%%% If preparing manually:
% \begin{thebibliography}{1}
% \newcommand{\enquote}[1]{``#1''}

% \bibitem{Zhang:14}
% Y.~Zhang, S.~Qiao, L.~Sun, Q.~W. Shi, W.~Huang, L.~Li, and Z.~Yang,
%   \enquote{Photoinduced active terahertz metamaterials with nanostructured
%   vanadium dioxide film deposited by sol-gel method,}
%   {\protect\JournalTitle{Optics Express}} \textbf{22}, 11070--11078 (2014).

% \bibitem{Optica}
% {Optica}, \enquote{{Optica Publishing Group},}
%   \url{http://www.opg.optica.org}.

% \bibitem{FORSTER2007}
% P.~Forster, V.~Ramaswamy, P.~Artaxo, T.~Bernsten, R.~Betts, D.~Fahey,
%   J.~Haywood, J.~Lean, D.~Lowe, G.~Myhre, J.~Nganga, R.~Prinn, G.~Raga,
%   M.~Schulz, and R.~V. Dorland, \enquote{Changes in atmospheric consituents and
%   in radiative forcing,} in \enquote{Climate Change 2007: The Physical Science
%   Basis. Contribution of Working Group 1 to the Fourth assesment report of
%   Intergovernmental Panel on Climate Change,}  S.~Solomon, D.~Qin, M.~Manning,
%   Z.~Chen, M.~Marquis, K.~B. Averyt, M.~Tignor, and H.~L. Miler, eds.
%   (Cambridge University Press, 2007).

% \end{thebibliography}

\end{document}